\documentclass[aps,
			   twocolumn,
			   nofootinbib,
			   amsmath,
			   amssymb,
			  ]{revtex4}

\usepackage[utf8]{inputenc}
\usepackage{combelow} 
\usepackage{microtype}
\usepackage{graphicx}

\usepackage{natbib} 

\usepackage{color}
\usepackage{url}
\usepackage{hyperref}
\hypersetup{
  colorlinks   = true,    
  citecolor    = black,   
  filecolor    = black, 
  linkcolor    = black,   
  urlcolor     = black    
}

\usepackage{amssymb} 
\usepackage{amsmath} 
\usepackage{amsfonts}
\usepackage{bm}
\usepackage{latexsym}%

\usepackage{subfig}

\usepackage{multirow}
\usepackage{graphicx}
\usepackage{lscape}
\usepackage{booktabs}
\usepackage{textcomp}
\usepackage{array}
\usepackage{makecell}

\begin{document}

	\title{Dataset Augmentation and Dimensionality Reduction of Pinna-Related Transfer Functions}

	\author{\href{https://orcid.org/0000-0002-9823-335X}{\includegraphics[scale=0.06]{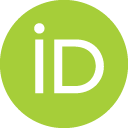}\hspace{1mm}Corentin \surname{Guezenoc}}}
	\email{corentin.guezenoc@centralesupelec.fr}
	\affiliation{FAST Research Team \\IETR (CNRS UMR 6164) \\CentraleSup\'elec \\Rennes, France}

	\author{\href{https://orcid.org/0000-0001-7199-7563}{\includegraphics[scale=0.06]{orcid.png}\hspace{1mm}Renaud \surname{S\'eguier}}}
	\email{renaud.seguier@centralesupelec.fr}
	\affiliation{FAST Research Team \\IETR (CNRS UMR 6164) \\CentraleSup\'elec \\Rennes, France}

	\begin{abstract}
		Efficient modeling of the inter-individual variations of head-related transfer functions (HRTF) is a key matter to the individualization of binaural synthesis. 
In previous work, we augmented a dataset of 119 pairs of ear shapes and pinna-related transfer functions (PRTFs), thus creating a wide dataset of 1005 ear shapes and PRTFs generated by random ear drawings (WiDESPREaD) and acoustical simulations. 
In this article, we investigate the dimensionality reduction capacity of two principal component analysis (PCA) models of magnitude PRTFs, trained on WiDESPREaD and on the original dataset, respectively. 
We find that the model trained on the WiDESPREaD dataset performs best, regardless of the number of retained principal components. 
	\end{abstract}

	\maketitle

\section{Introduction}
Head-related transfer functions (HRTFs) individualization is a key matter in binaural synthesis \cite{wenzel_localization_1993}. 
A lot of work has been done towards proposing user-friendly personalization methods, either based on anthropometric measurements \cite{middlebrooks_individual_1999, zotkin_customizable_2002, hu_head_2006} or on perceptual feed-back \cite{middlebrooks_psychophysical_2000, seeber_subjective_2003, hwang_modeling_2008-1, shin_enhanced_2008, yamamoto_fully_2017}, most of which rely on HRTF databases. 
Although some of these approaches propose to select a best-mach HRTF set among a database  \cite{seeber_subjective_2003}, others rely on a statistical model of the inter-individual variations of HRTF sets  \cite{hu_head_2006, hwang_modeling_2008-1, shin_enhanced_2008, yamamoto_fully_2017}.
In particular, an interesting direction of work consists in fitting an HRTF model to the listener based on his perceptual feed-back \cite{hwang_modeling_2008-1, shin_enhanced_2008, yamamoto_fully_2017}, either by tuning himself the model parameters, or by using an optimization process that prompts the listener for perceptual feed-back.
In this context, tuning time must be kept as low as possible. 
Thus, it is essential for the HRTF model to be as compact as possible, i.e. having as few parameters as possible. 

However, the databases that are currently available have few subjects compared to the dimensionality of the data.
Indeed, the largest one, the Acoustics Research Institute (ARI) dataset \cite{majdak_3-d_2010},  includes HRTF sets for 201 subjects,  while the dimensionality of the data is of roughly $1.2 \cdot 10^6$ (256 time-domain samples $\times$ 2300 directions $\times$ 2 ears) for a typical high-resolution HRTF set \cite{bomhardt_high-resolution_2016}.
To address this matter, in previous work \cite{guezenoc_wide_2020} we proposed a method to augment a dual dataset of 119 ear point clouds \cite{ghorbal_method_2019} and corresponding pinna-related transfer functions (PRTFs).
The method consists in drawing new ear shapes according to the observed statistical distribution then computing the corresponding PRTF sets by means of fast-multipole boundary element method (FM-BEM).
The resulting augmented dataset, named WIDESPREaD (wide dataset of ear shapes and pinna-related transfer functions generated by random ear drawings) includes 1005 artificial subjects and is freely available on the sofacoustics website\footnote{\url{https://www.sofacoustics.org/data/database/widespread}}.

In the present article, we look into how using this new dataset instead of the original one for training improves the capacity of principal component analysis (PCA) to reduce the dimensionality of log-magnitude\footnote{We focus here on the magnitude spectra, as the perceptual defects due to a lack of individualization mostly derive from the distortion of the spectral cues \cite{asano_role_1990}.} PRTF sets.
To this end, a 20-fold cross-validation of PCA is performed on each dataset.
The training and validation reconstruction errors are then compared for various numbers of retained principal components (PCs).


The paper is organized as follows.
First, we present the original dataset of 119 pairs of ear shape and corresponding simulated PRTF sets.
Second, we summarize the process of data augmentation that led to the creation of the WiDESPREaD dataset.
Third, we present the construction of two PCA models of magnitude PRTF set, trained on the original and augmented dataset, respectively.
Fourth, we discuss the dimensionality reduction capacity of both PCA models by studying the cumulative proportion of variance, then by performing a 20-fold cross-validation in order to evaluate their performance on data left out of the training set.
Finally, we discuss the results.

\section{A Dataset of 3-D Ear Scans and Matching Simulated PRTFs} 
\label{sec:OriginalDataset}

The original dataset is composed of pairs of ear mesh and matching simulated PRTF set for 119 human subjects. 

The database of 119 ear meshes was constituted in previous work by Ghorbal \emph{et al.} \cite{ghorbal_method_2019}.
As part of that work, 3D scans were acquired using a structure-light based scanner, before being normalized in size and rigidly aligned.
Finally, the point clouds were registered: every point cloud has the same number ${n_v} = 18176$ of vertices, and the vertex indexing is semantically coherent from one subject to the other.
Meshes can be derived from the point clouds thanks to a set of 35750 triangular faces defined identically for all point clouds by the indices of the ${n_v}$ vertices.

In order to compute the PRTF sets, we closed the ear meshes using a cylinder-like support mesh.
We then performed the acoustical simulations by FM-BEM using the mesh2hrtf \cite{ziegelwanger_mesh2hrtf:_2015} software by Ziegelwanger \emph{et al}.
This step is described in more details in \cite{guezenoc_wide_2020}.

\section{Dataset Augmentation} 

In a previous paper \cite{guezenoc_wide_2020}, exploiting the initial dataset of 119 pairs of ear shapes and PRTFs, we proposed a method to generate new examples based on a generative model of 3-D ear shapes and boundary element method simulations.
As a result, the WiDESPREaD dataset was created and made available to other researchers,  featuring 1005 artificial subjects.
In this section, we summarize the data augmentation process (the curious reader can refer to \cite{guezenoc_wide_2020} for more details).

\subsection{Statistical Ear Shape Model} 
\label{subsec:StatisticalEarShapeModel}

Thanks to the fact that the ear point clouds are in correspondence, we were able to perform a PCA.

By concatenating the $x$, $y$ and $z$ coordinates, every point cloud was represented as a row vector $\mathbf{e}_i$ of $\mathbb{R}^{3 {n_v}}$, where $i = 1, \; \dots \; {N_\mathrm{O}}$, with ${N_\mathrm{O}}$ the number of subjects in the original dataset.
Let us denote by $\mathrm{E} = \left\lbrace \mathbf{e}_1, \mathbf{e}_2, \; \dots \; \mathbf{e}_{N_\mathrm{O}} \right\rbrace$ the dataset of 119 ear point clouds.
The corresponding data matrix $\mathbf{X}_\mathrm{E} \in \mathbb{R}^{{N_\mathrm{O}} \times 3 {n_v}}$ was constructed by stacking the ear point cloud row vectors vertically 
$\mathbf{X}_\mathrm{E} =
{ \left( {\mathbf{e}_1}^\mathrm{t}, \; \dots \; {\mathbf{e}_{N_\mathrm{O}}}^\mathrm{t} \right) }^\mathrm{t}$.
Additionally, let $\bar{\mathbf{e}} = \frac{1}{N_\mathrm{O}}  \displaystyle\sum_{i=1}^{N_\mathrm{O}} \mathbf{e}_i$ 
be the average ear shape
and
$\bar{\mathbf{X}}_\mathrm{E} = { \left( \bar{\mathbf{e}} \; \dots \; \bar{\mathbf{e}} \right) }^\mathrm{t} \in \mathbb{R}^{{N_\mathrm{O}} \times {3 {n_v}}}$ the matrix constituted of the average shape stacked ${N_\mathrm{O}}$ times, 
and let $\bm{\Gamma}_\mathrm{E} \in \mathbb{R}^{3 {n_v} \times 3 {n_v}}$ be the covariance matrix of $\mathbf{X}_\mathrm{E}$:
\begin{equation}
	\bm{\Gamma}_\mathrm{E} = 
	\frac{1}{{N_\mathrm{O}} - 1} 
	{\left( \mathbf{X}_\mathrm{E} - \bar{\mathbf{X}}_\mathrm{E} \right)}^\mathrm{t} 
	\left( \mathbf{X}_\mathrm{E} - \bar{\mathbf{X}}_\mathrm{E} \right) 
	.
\end{equation}

PCA can thus be written as
\begin{equation} 
\label{eq:earPca1}
	\mathbf{Y}_\mathrm{E} = \left( \mathbf{X}_\mathrm{E} - \bar{\mathbf{X}}_\mathrm{E} \right) {\mathbf{U}_\mathrm{E}}^\mathrm{t} ,
\end{equation}
where $\mathbf{U}_\mathrm{E}$ is obtained by diagonalizing $\bm{\Gamma}_\mathrm{E}$
\begin{equation}
	\bm{\Gamma}_\mathrm{E} = 
	\mathbf{U}_\mathrm{E}^\mathrm{t} 
	{\bm{\Sigma}_\mathrm{E}}^2 
	{\mathbf{U}_\mathrm{E}}   
	 ,
\end{equation}
with ${\bm{\Sigma}_\mathrm{E}}^2 \in \mathbb{R}^{({N_\mathrm{O}}-1) \times ({N_\mathrm{O}}-1)}$ a diagonal matrix that contains its eigenvalues ${\sigma_{\mathrm{E}_1}}^2 , \, {\sigma_{\mathrm{E}_2}}^2 , \, \dots {\sigma_{\mathrm{E}_{{N_\mathrm{O}} - 1}}}^2$ 
\begin{equation}
\label{eq:earPca2}
	{\bm{\Sigma}_\mathrm{E}}^2 = 
	\begin{bmatrix}
		{\sigma_{\mathrm{E}_1}}^2 & & \\
		 & \ddots & \\
		& & {\sigma_{\mathrm{E}_{{N_\mathrm{O}} - 1}}}^2 \\
	\end{bmatrix}  
\end{equation}
ordered so that ${\sigma_{\mathrm{E}_1}}^2 \geq {\sigma_{\mathrm{E}_2}}^2 \geq \dots \geq {\sigma_{\mathrm{E}_{{N_\mathrm{O}} - 1}}}^2$,
and with
$\mathbf{U}_\mathrm{E} \in \mathbb{R}^{({N_\mathrm{O}} - 1) \times 3 {n_v}}$ an orthogonal matrix that contains the corresponding eigenvectors ${\mathbf{u}_{\mathrm{E}_1}} , \, {\mathbf{u}_{\mathrm{E}_2}}, \; \dots \; {\mathbf{u}_{\mathrm{E}_{{N_\mathrm{O}}-1}}} \in \mathbb{R}^{3 {n_v}}$ 
\begin{equation}
	\mathbf{U}_\mathrm{E} = 
	\begin{bmatrix}
		{\mathbf{u}_{\mathrm{E}_1}} \\
		\vdots \\
		{\mathbf{u}_{\mathrm{E}_{{N_\mathrm{O}}-1}}}
	\end{bmatrix}
	. 
\end{equation}
The eigenvalues denote how much variance in the input data is explained by the corresponding eigenvectors.

\subsection{Random Drawing of Ear Shapes}

The PCA model was then used as a generative model to randomly draw an arbitrary large number ${N_\mathrm{W}}$ of new ear shapes.

For all subject of index $j = 1, \; \dots \; {N_\mathrm{W}}$,
a principal component (PC) weights vector $\mathbf{y}_{{\mathrm{E}_\mathrm{W}}_j} = (y_{{\mathrm{E}_\mathrm{W}}_{j, 1}}, \; \dots \; y_{{\mathrm{E}_\mathrm{W}}_{j, {N_\mathrm{O}} - 1}} ) \in \mathbb{R}^{{N_\mathrm{O}} - 1}$
was obtained by drawing the $({N_\mathrm{O}} - 1)$ PC weights $y_{{\mathrm{E}_\mathrm{W}}_{j, 1}}, \; \dots \; y_{{\mathrm{E}_\mathrm{W}}_{j, {N_\mathrm{O}} - 1}}$ independently according to their respective observed probability laws $\mathcal{N}(0, \sigma_{\mathrm{E}_1}^2), \, \dots \mathcal{N}(0, \sigma_{\mathrm{E}_{{N_\mathrm{O}}-1}}^2)$.

Then, 
the corresponding ear shapes were reconstructed by inverting Equation~\eqref{eq:earPca1}
\begin{equation}
 	\begin{bmatrix}
		\mathbf{e_\mathrm{W}}_1 \\
		\vdots \\
		\mathbf{e_\mathrm{W}}_{N_\mathrm{W}} \\
	\end{bmatrix} = 
 	\mathbf{X}_{\mathrm{E}_\mathrm{W}} = 
	\mathbf{U}_\mathrm{E} \mathbf{Y}_{\mathrm{E}_\mathrm{W}} + \bar{\mathbf{X}}_\mathrm{E}
	,
\end{equation}
where $\mathbf{Y}_{\mathrm{E}_\mathrm{W}} \in \mathbb{R}^{{N_\mathrm{W}} \times ({N_\mathrm{O}} - 1)}$ is the matrix whose rows are the ${N_\mathrm{W}}$ PC weights vectors
\begin{equation}
	\mathbf{Y}_{{\mathrm{E}_\mathrm{W}}} = 
	\begin{bmatrix}
		 \mathbf{y}_{{\mathrm{E}_\mathrm{W}}_1} \\
		\vdots \\
		\mathbf{y}_{{\mathrm{E}_\mathrm{W}}_{{N_\mathrm{W}}}} \\
	\end{bmatrix}
	=
	\begin{bmatrix}
		y_{{\mathrm{E}_\mathrm{W}}_{1, 1}} \! \! \! \! \! & \dots \! \! \! \! \! & y_{{\mathrm{E}_\mathrm{W}}_{1, {N_\mathrm{O}} - 1}} \\
		\vdots \! \! \! \! \! & \ddots \! \! \! \! \! & \vdots \\
		y_{{\mathrm{E}_\mathrm{W}}_{{N_\mathrm{W}}, 1}} \! \! \! \! \! & \dots \! \! \! \! \! & y_{{\mathrm{E}_\mathrm{W}}_{{N_\mathrm{W}}, {N_\mathrm{O}} - 1}}  \\
	\end{bmatrix}
	.
\end{equation}

\subsection{Simulation of PRTF Sets}

Finally, ${N_\mathrm{W}}$ PRTF sets were calculated from the ${N_\mathrm{W}}$ ear point clouds as it was done in the case of the original dataset (see Sub-Section~\ref{sec:OriginalDataset}).

\begin{figure*}
    \subfloat[]{
    	\includegraphics[width=0.98\textwidth]{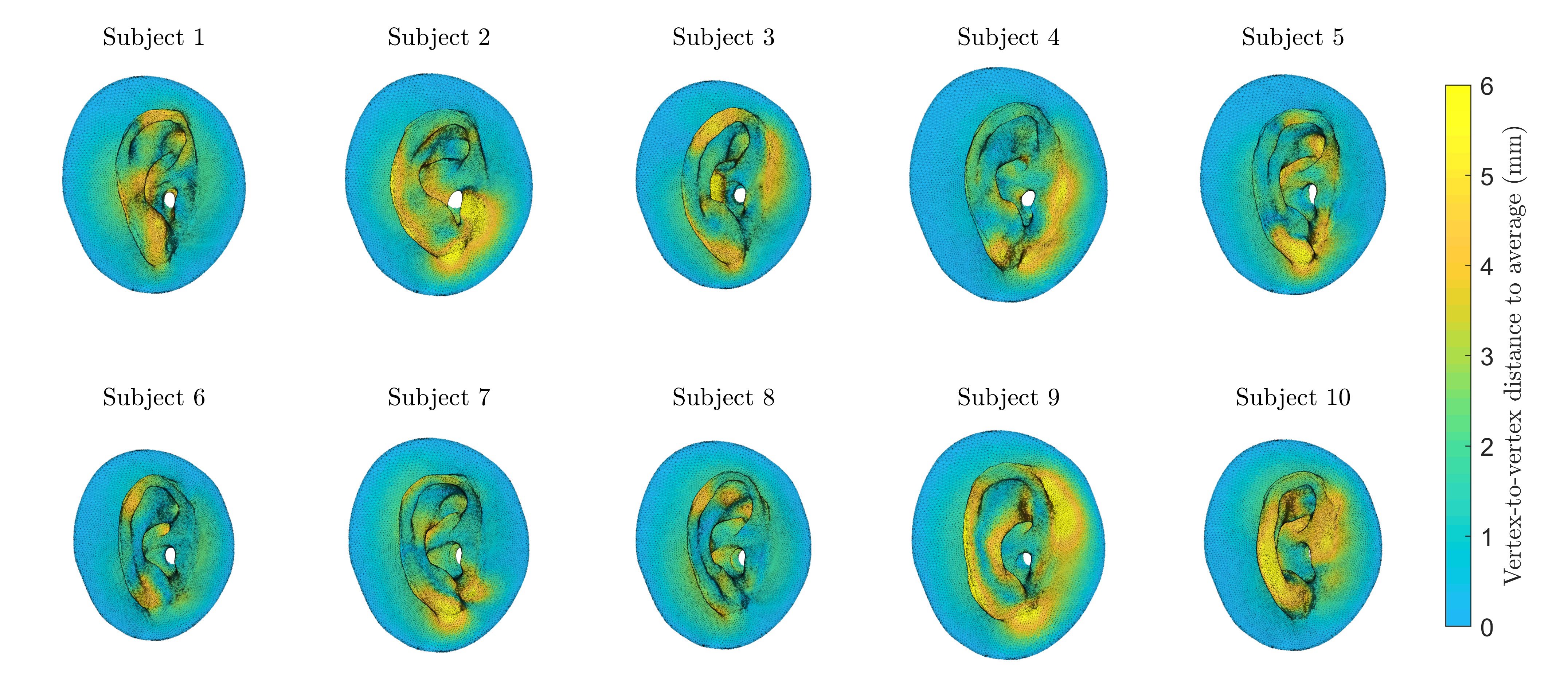}
    	}
	
    \subfloat[]{
    	\includegraphics[width=0.98\textwidth]{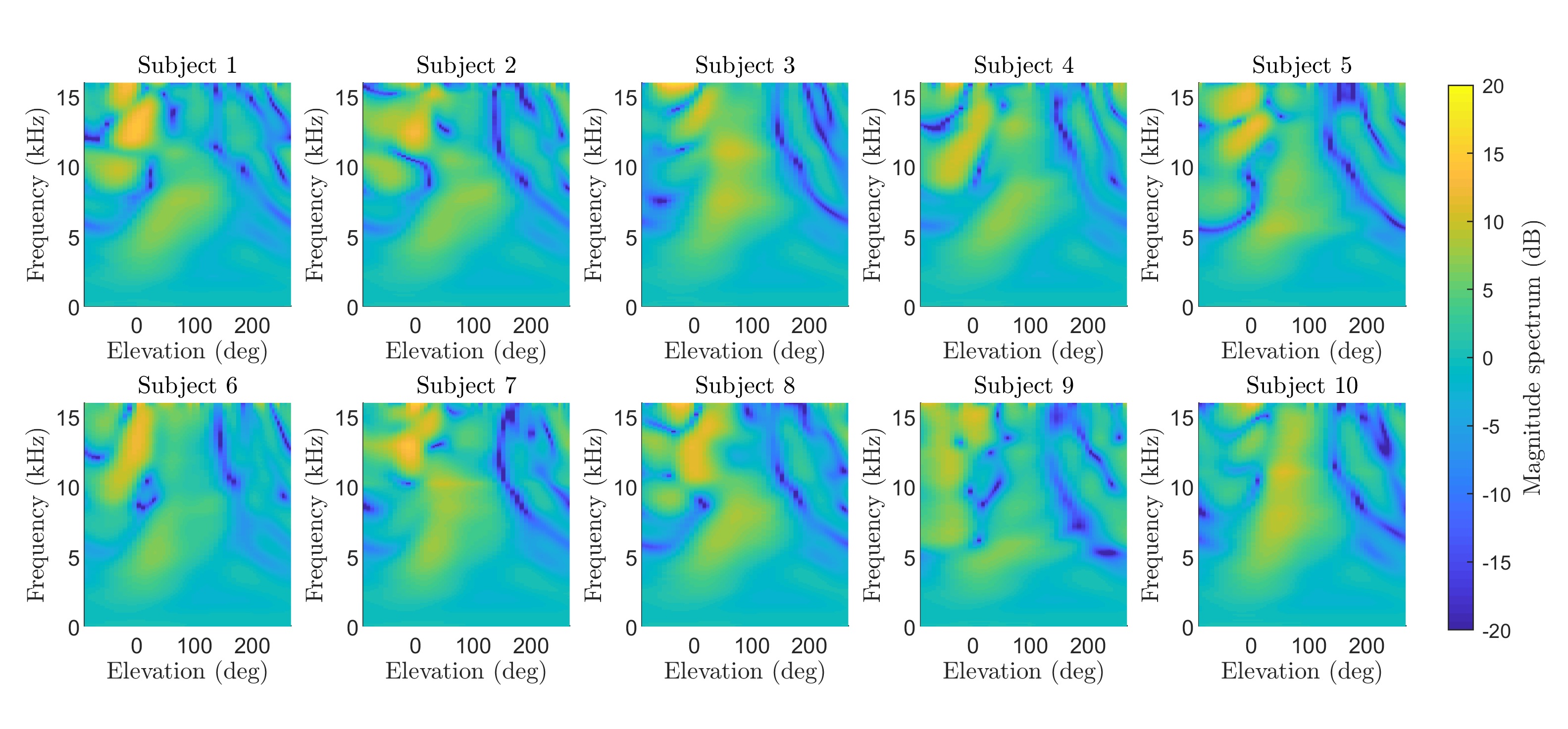}
    	}

\caption{
Visualization of the first 10 subjects of WiDESPREaD.
(a) Meshes derived from the synthetic ear shapes $\mathbf{e}_{\mathrm{W}_1}, \; \dots \; \mathbf{e}_{\mathrm{W}_{10}}$. Color represents the vertex-to-vertex euclidean distance to the generative model's average $\bar{\mathbf{e}}$. 
(b) Log-magnitude PRTF sets $20 \cdot \log_{10} ( \mathbf{p}_{\mathrm{W}_1} ), \; \dots \; 20 \cdot \log_{10} ( \mathbf{p}_{\mathrm{W}_{10}} )$ 
displayed in the median sagittal plane.
}
\label{fig:WidespreadExamples}
\end{figure*}

For illustration purposes, the ear shapes and PRTF sets of the first 10 artificial subjects of WiDESPREaD are displayed in Figure~\ref{fig:WidespreadExamples}.

\section{Principal Component Analysis of Magnitude PRTFs}

Let us denote by $\mathbf{p}_{\mathrm{O}_1}, \; \dots \; \mathbf{p}_{\mathrm{O}_{N_\mathrm{O}}} \in \mathbb{R}^{n_f \times n_d}$ and $\mathbf{p}_{\mathrm{W}_1}, \; \dots \; \mathbf{p}_{\mathrm{W}_{N_\mathrm{W}}} \in \mathbb{R}^{n_f \times n_d}$ the log-magnitude PRTF sets of the original and WiDESPREaD datasets, respectively,
where log-magnitude PRTFs are meant here as $20 \cdot \log_{10} \left( | \cdot | \right)$ of the complex PRTFs.

We concatenated the filters from all $n_d$ directions of each log-magnitude PRTF set $\mathbf{p}_{S_i} \in \mathbb{R}^{n_f \times n_d}$ into a row vector of $\mathbb{R}^{n_f n_d}$,  where $i = 1, \; \dots \; {N_S}$ and $S \in \lbrace \mathrm{O}, \, \mathrm{W} \rbrace$. 
Further on, $\mathbf{p}_{S_i}$ designates the row vector representation of the log-magnitude PRTF set.
The logarithmic scale was chosen for its coherence with human perception.


The data matrices $\mathbf{X}_\mathrm{O} \in \mathbb{R}^{{N_\mathrm{O}} \times n_f n_v}$ and $\mathbf{X}_\mathrm{W} \in \mathbb{R}^{{N_S} \times n_f n_v}$ were then constituted by stacking vertically the log-magnitude PRTF sets:
$
	\mathbf{X}_\mathrm{O} = 
	\left( 
	{\mathbf{p}_{\mathrm{O}_1}}^\mathrm{t} 
	\; \dots \; 
	{\mathbf{p}_{\mathrm{O}_{N_\mathrm{O}}}}^\mathrm{t}
	\right)^\mathrm{t}
$
 and 
$
	\mathbf{X}_\mathrm{W} = 
	\left(
		{\mathbf{p}_{\mathrm{W}_1}}^\mathrm{t}
		\; \dots \; 
		{\mathbf{p}_{\mathrm{W}_{N_\mathrm{W}}}}^\mathrm{t}
	\right)^\mathrm{t}
	.
$
PCA was then performed on each dataset as described by Equations~\eqref{eq:earPca1} and \eqref{eq:earPca2} in the case of ear point clouds.

\section{Dimensionality Reduction Capacity}

PCA can be used as a dimensionality reduction technique by only retaining the first ${m}$ PCs \cite{jolliffe_principal_2002-1}, where ${m} \in \lbrace 1, \; \dots \; {N_S}-1 \rbrace$ and $S \in \left\lbrace \mathrm{O}, \, \mathrm{W} \right\rbrace$ designates the dataset:
\begin{equation} 
\label{eq:pcaDimRed}
	 \tilde{\mathbf{Y}}_S^{(m)} = 
	\begin{bmatrix}
		y_{S_{1, 1}} & \dots   & y_{S_{1, {m}}} & 0          & \dots   & 0 \\
		\vdots  & \ddots & \vdots  &  \vdots & \ddots & \vdots \\
		y_{S_{{N_S}, 1}} & \dots   &y_{S_{{N_S}, {m}}} & 0           & \dots   & 0 \\
	\end{bmatrix}
	,
\end{equation}
where $y_{S_{i, j}}$ is the value of matrix $\mathbf{Y}_S$ at the $i^\mathrm{th}$ row and $j^\mathrm{th}$ column for all $i = 1, \; \dots \; {N_S}$ and $j = 1, \; \dots \; {{N_S} - 1}$.

\subsection{Cumulative Percentage of Total Variance}
\label{subsec:Cpv}

Approximated data can be then reconstructed by inverting Equation~\eqref{eq:earPca1}:
\begin{equation}
	\label{eq:ReconsApprox}
	\tilde{\mathbf{X}}_S^{(m)}  =  \tilde{\mathbf{Y}}_S^{(m)} {\mathbf{U}_S} + \bar{\mathbf{X}}_S
	.
\end{equation}

A simple but useful metric to evaluate the dimensionality reduction capacity of a PCA model is the cumulative percentage of total variance (CPV) \citep[section 6.1]{jolliffe_principal_2002-1}:
\begin{equation}
{\mathrm{CPV}_S}({m}) = 
100 \cdot 
\left( \displaystyle \sum_{j=1}^{m} {{\sigma_S}_j}^2 \right)  
/ 
\left( {\displaystyle\sum_{j=1}^{{N_S} - 1} {{\sigma_S}_j}^2} \right)
, \;
\end{equation}
where $S \in \left\lbrace \mathrm{O}, \mathrm{W} \right\rbrace$ represents either the original or the WiDESPREaD set of log-magnitude PRTFs
and where ${m} \in \lbrace 1 , \; \dots \; {N_S} - 1 \rbrace$ is the number of retained PCs. 

Let us note that CPV is closely related to the dimensionality reduction-related mean-square reconstruction error (MSE) of the training set \citep[section 6.1]{jolliffe_principal_2002-1}. 
This relation can be expressed as follows: 
\begin{equation}
	\label{eq:CpvVsDimRedError}
	{\mathrm{CPV}_S}({m}) = 100 \cdot 
	\frac 
		{1 -  \mathrm{MSE} ( \tilde{\mathbf{X}}_S^{(m)} ,  {\mathbf{X}}_S)}
		{\mathrm{MSE} (\bar{\mathbf{X}}_S, \mathbf{X}_S)} 
		,
\end{equation}
where 
\begin{equation}
	\label{eq:mse}
\mathrm{MSE} ( \mathbf{A} ,  \mathbf{B} ) = 
  \frac{1}{q} \frac{1}{r}
  \left( \mathbf{A}  - \mathbf{B} \right) \left( \mathbf{A}  - \mathbf{B}  \right)^\mathrm{t}, 
\end{equation}
  for all $\mathbf{A}, \mathbf{B} \in \mathbb{R}^{q \times r}$ and $q, r \in \mathbb{N}$
  .

\begin{figure}
	\includegraphics[width=0.47\textwidth]{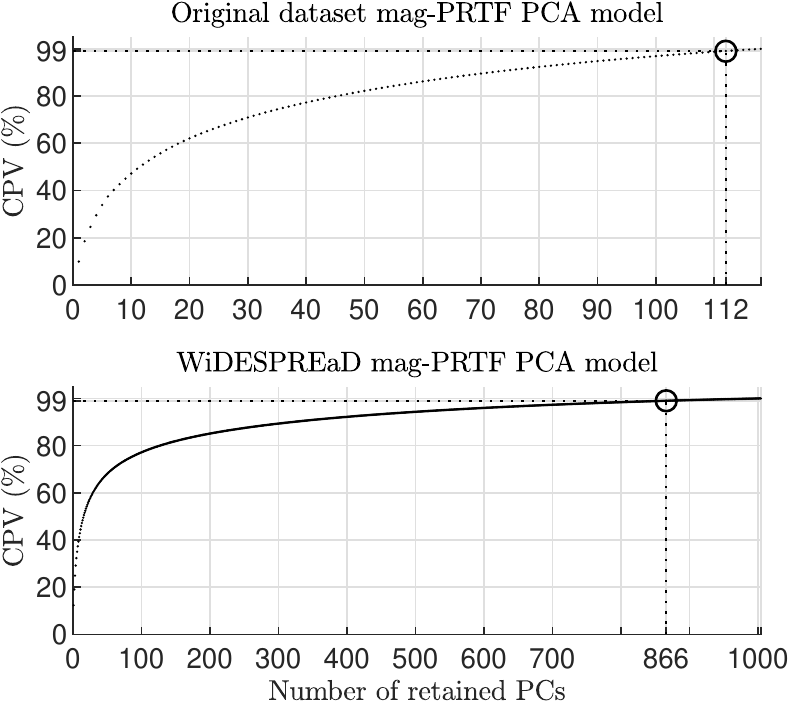}
	\caption{Cumulative percentage of total variance (CPV) of log-magnitude PRTF PCA models as a function of number of retained principal components.
				Please note that the x-axes are on different scales, due to the large difference in number of subjects (and thus number of PCs) between models.
				Up: Original dataset.
				Down: WiDESPREaD.}
	\label{fig:VarExpPrtfNewVsOld}
\end{figure}

CPVs of both original and WiDESPREaD log-magnitude PRTF PCA models are plotted in Figure~\ref{fig:VarExpPrtfNewVsOld}.
%
In the case of the original dataset, we observe, as in \cite{guezenoc_wide_2020}, that most PCs i.e. $95 \%$ (112 out of 118) of them are required to reach the threshold of $99 \%$ of the total variance. 
It would seem that we can not find a linear sub-space of the space generated by our ${N_\mathrm{O}} = 119$ examples in which log-magnitude PRTF sets are well represented.
As far as we can see, a first possible explanation is that log-magnitude PRTF sets populate a linear sub-space of $\mathbb{R}^{n_f n_d}$ whose dimension is greater than ${N_\mathrm{O}} - 1 = 118$. 
A second explanation may be the existence of a non-linear manifold which would cause PCA to require all principal components to describe accurately the training set, independently of the number of training examples.
Either way, it is hard to conclude without more subjects, which initially motivated the creation of the WiDESPREaD dataset.

It can be observed that, for equal numbers of retained PCs, the CPV is lower in the original case than in the WiDESPREaD one.
For instance, a CPV of $99 \%$ is reached using 112 PCs for the first while many more, i.e. 866, are needed for the latter. 
Although this may appear as a regression, let us bear in mind that the CPV represents a proportion of the total variance observed in the dataset.
However, the total variance is not the same in both datasets.
Indeed, the examples of magnitude PRTFs in WiDESPREaD are more numerous and were synthesized so that their statistical distribution was realistic.
Hence, as expected, the examples of this dataset are more diverse than that of the original dataset, resulting in a higher total variance.
In other words, the new PCA model captures variations of magnitude PRTFs that were not present in the original dataset, and naturally uses more principal components to do so.

In the case of WiDESPREaD, the $99 \%$ CPV threshold is reached by retaining $86 \%$ (866 out of 1004) of the PCs.
In other words, the last 138 principal components are not needed to approximate log-magnitude PRTF sets of the training set with reasonable accuracy.
The ratio of PCs to be retained is noticeably smaller than in the case of the original dataset, which is an indication that the model trained on WiDESPREaD PRTFs may perform better at representing log-magnitude PRTF sets in general.


\subsection{20-Fold Cross-Validation}
\label{subsec:CrossValidation}
In order to assess and compare the capacity of both PCA models to generalize to new examples, we performed a 20-fold cross-validation for each one of them.

Each dataset $S \in \left\lbrace \mathrm{O}, \, \mathrm{W} \right\rbrace$ was equally divided in $K=20$ sub-groups, each containing about $5 \%$ of the subjects.
Each sub-group of index $k = 1, \; \dots \; K$ was then used in turn as a validation set for a PCA model trained the subjects of the remaining $K-1$ folds.
For all ${k = 1, \; \dots \; K}$ and for all dataset $S \in \left\lbrace \mathrm{O}, \, \mathrm{W} \right\rbrace$, let there be $I_{S_{\mathrm{train}, k}} \subset \left\lbrace 1, \, \dots, {N_S} \right\rbrace$ and $I_{S_{\mathrm{val}, k}} \subset \left\lbrace 1, \, \dots, {N_S} \right\rbrace$ the sets of subject indices that constitute the $k^\mathrm{th}$ fold's training and validation sets, respectively.

Let there be a fold $k = 1, \; \dots \; K$ and a dataset ${S \in \left\lbrace \mathrm{O}, \, \mathrm{W} \right\rbrace}$. 
PCA was performed on the data matrix $\mathbf{X}_{S_{\mathrm{train}, k}} = \left( \mathbf{p}_{S_i} \right)_{i \in I_{S_{\mathrm{train}, k}}}$. 
Re-writing Equation~\eqref{eq:earPca1} using this notation, the PCA transform can be written:
\begin{equation}
	\label{eq:PcaCrossVal}
	\mathbf{Y}_{{S_{\mathrm{train}, k}}} = \left( \mathbf{X}_{{S_{\mathrm{train}, k}}} - \bar{\mathbf{X}}_{S_{\mathrm{train}, k}} \right) {\mathbf{U}_{{S_{\mathrm{train}, k}}}}^\mathrm{t} 
	.
\end{equation}

Examples from the validation set $\mathbf{X}_{S_{\mathrm{val}, k}} = \left( \mathbf{p}_{S_i} \right)_{i \in I_{S_{\mathrm{val}, k}}}$ were then projected in the training space as follows:
\begin{equation}
	\label{eq:ProjectVali}
	\mathbf{Y}_{{S_{\mathrm{val}, k}}} = \left( \mathbf{X}_{{S_{\mathrm{val}, k}}} - \bar{\mathbf{X}}_{S_{\mathrm{train}, k}} \right) {\mathbf{U}_{{S_{\mathrm{train}, k}}}}^\mathrm{t} 
	.
\end{equation}

Finally, the training and validation data matrices were reconstructed from the PC weights.
The number of PCs retained for reconstruction, $m$, varied in $\lbrace 0, \; \dots \; {\mathrm{card}(I_{S_{\mathrm{val}, k}}) - 1} \rbrace$, where $\mathrm{card}(I_{S_{\mathrm{train}, k}}) = (K - 1) \left\lfloor \frac{{N_S}}{K} \right\rfloor$ is the number of training subjects. 
Thus, using the same notation as in Equation~\eqref{eq:pcaDimRed} and according to \eqref{eq:ReconsApprox}, training and validation sets were reconstructed according to the following equations:
\begin{equation}
	\tilde{\mathbf{X}}_{{S_{\mathrm{train}, k}}}^{(m)} =  \tilde{\mathbf{Y}}_{S_{\mathrm{train}, k}}^{(m)} {\mathbf{U}_{S_{\mathrm{train}, k}}} + \bar{\mathbf{X}}_{S_{\mathrm{train}, k}}
	,
\end{equation}
and
\begin{equation}
	\tilde{\mathbf{X}}_{{S_{\mathrm{val}, k}}}^{(m)} =  \tilde{\mathbf{Y}}_{S_{\mathrm{val}, k}}^{(m)} {\mathbf{U}_{S_{\mathrm{train}, k}}} + \bar{\mathbf{X}}_{S_{\mathrm{train}, k}}
	.
\end{equation}

The MSE reconstruction error was then averaged across all folds for both training sets
\begin{equation}
	\epsilon_\mathrm{MSE, train}(S) = 
	\frac{1}{K} 
	\sum_{k = 1}^K 
	\mathrm{MSE} \left( \tilde{\mathbf{X}}_{{S_{\mathrm{train}, k}}}^{(m)}, {\mathbf{X}}_{{S_{\mathrm{train}, k}}} \right)
,
\end{equation}
and validation sets
\begin{equation}
	\epsilon_\mathrm{MSE, val}(S) = 
	\frac{1}{K} 
	\sum_{k = 1}^K 
	\mathrm{MSE} \left( \tilde{\mathbf{X}}_{{S_{\mathrm{val}, k}}}^{(m)}, {\mathbf{X}}_{{S_{\mathrm{train}, k}}} \right)
.
\end{equation}


\begin{figure*}
	\includegraphics[width=0.98\textwidth]{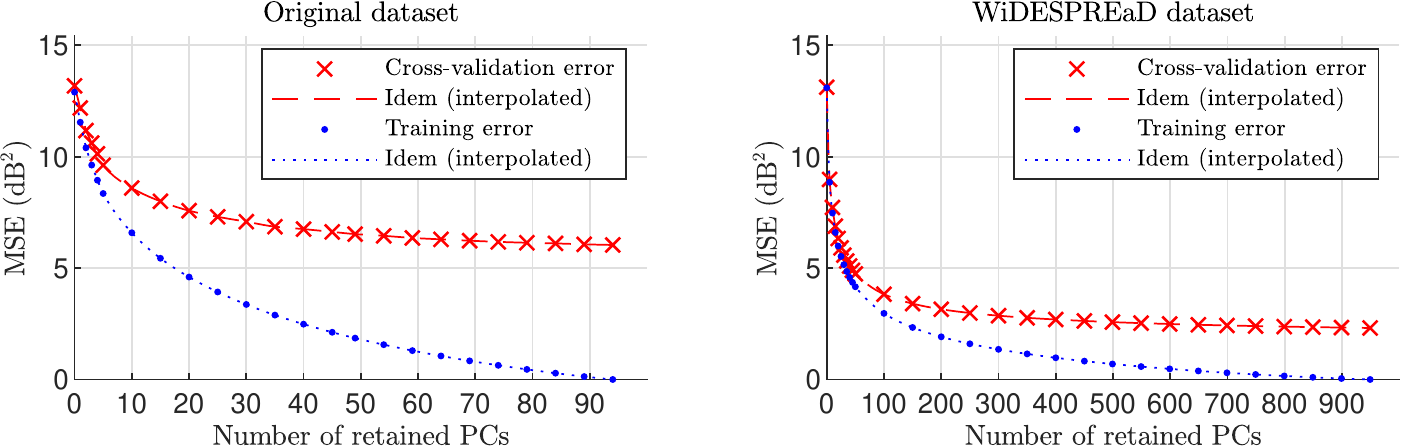}
	\caption{Mean-square reconstruction errors (MSE) of log-magnitude PRTF sets, averaged over 20 cross-validation folds, for various numbers of retained PCs, for both original (left) and WiDESPREaD (right) datasets.
					Training errors are displayed as blue points while validation errors are represented as red crosses.
					The blue dashed and red dotted lines are a cubic interpolation of the training and validation errors, respectively.
					Please note that the x-axes are on different scales, due to the large difference in number of subjects between models.
				    }
	\label{fig:CrossVal}
\end{figure*}

The training and validation reconstruction errors for both original and WiDESPREaD dataset are displayed in Figure~\ref{fig:CrossVal}.
A first observation is that, when retaining all principal components in both models, the WiDESPREaD PCA model reconstructs validation data with a notably lower error ($2.3~\mathrm{dB}^2$) than the original model ($6.0~\mathrm{dB}^2$), that is a factor of $2.6$.

More interestingly, for identical numbers of retained components, the WiDESPREaD cross-validation error is always lower than that of the original model. 
In particular, the lowest cross-validation error ever attained by the original model, $6.0~\mathrm{dB}^2$ (for $m = 94$ retained PCs) is reached by the WiDESPREaD one using as few as 35 retained PCs.
In other words, 35 PCs are sufficient to obtain a generalization error lower than when using the original model with all of its 94 PCs.

\section{Discussion}
\label{sec:discussion}
One of the observations that led to the creation of WiDESPREaD was the fact that, in the original log-magnitude PRTF PCA model, almost all PCs, i.e. $95 \%$ (112 / 118) of the PCs were necessary to retain at least $99 \%$ of the total variance observed in the dataset \cite{guezenoc_wide_2020}. 
Reassuringly, by training in the same manner a log-magnitude PRTF PCA model on the augmented PRTF dataset, we observe that a smaller ratio of PCs of $86 \%$ (866 / 1004) is required to reach a CPV of $99 \%$ (see Sub-Section~\ref{subsec:Cpv}).
Thus, it would seem that increasing the number of subjects allowed us indeed to identify a linear sub-space of $\mathbb{R}^{N_S - 1}$ that is able to contain most of the inter-individual variability of log-magnitude PRTF sets. 

Nevertheless, this is to be taken with caution. 
Indeed, while we increased the number of subjects by a factor 8.5, the decrease in the ratio of PCs required to reach a CPV of $99 \%$ is rather modest ($95 \%$ to  $86 \%$).
Hence, the possibility remains that adding more subjects would keep increasing significantly the number of PCs required to represent $99 \%$ the information, which would be the case if there was a non-linear manifold, for instance.

However, cross-validation yielded promising results as to providing a compact representation of log-magnitude PRTF sets in general (see Sub-Section~\ref{subsec:CrossValidation}).
Indeed, the PCA model based on WiDESPREaD seemed able to generalize to new examples better (lower reconstruction error) than the model trained on the original dataset.

When comparing both models with all their PCs retained, it could be expected.
Indeed, approximating new data thanks to a PCA model while retaining all PCs is equivalent to a projection into the $(N_S-1)$-dimensional space generated by linear combinations of the $N_S$ training examples,
and the WiDESPREaD dataset has about 9 times more examples than the original one.

However, and more interestingly, the generalization error curve for the WiDESPREaD case is lower than that of the original dataset, regardless of the number of retained PCs.

Furthermore, only 35 components (out of 949) are needed for the WiDESPREaD cross-validation reconstruction error to subceed the lowest cross-validation error ever attained in the case of the original dataset i.e. with 94 retained PCs out of 94. 

\section{Conclusion}
In this article, we presented a dataset of 119 pairs of ear mesh and matching calculated PRTF set.
Thereupon, we presented briefly how this dataset was augmented in previous work \cite{guezenoc_wide_2020}, resulting in a wide dataset of 1005 pairs of ear mesh and matching calculated PRTF set, named WiDESPREaD.
Building upon this original dataset, we then summarized how an augmented dataset of 1005 pairs of ear mesh and matching calculated PRTF set was generated, according to a method introduced in a previous article.
We then trained PCA models of log-magnitude PRTF sets on each original and augmented dataset,
before comparing their dimensionality reduction performance on both training and validation data using tools such as the cumulative proportion of total variance and k-fold cross-validation.

Overall, we found that using the WiDESPREaD dataset improved the performances of PCA at modeling and reducing the dimensionality of log-magnitude PRTF sets, in comparison with the original dataset.
These results are encouraging and tend to corroborate the promising character of the process of synthetic data generation that we proposed in \cite{guezenoc_wide_2020}, 
which could benefit to applications such as HRTF adaptation based on listener feed-back or anthropometry. 

As the possibility for a non-linear manifold in the space of log-magnitude PRTFs is not ruled out (see Section~\ref{sec:discussion}), future work includes using a non-linear dimensionality reduction technique on the database of WiDESPREaD PRTFs and compare its performance with PCA.

Furthermore, it may be interesting to use other metrics in addition to MSE to evaluate dimensionality reduction error, such as inter-subject spectral difference (ISSD) \cite{middlebrooks_individual_1999} and simulated sagittal plane sound localization errors using the psycho-acoustic model by Baumgartner \emph{et al.} \cite{baumgartner_modeling_2014}.



	\bibliography{zotero}

\end{document}